\documentstyle[PASJadd]{PASJ95}
\draft

\newcommand{\vol}{}
\newcommand{\no}{}
\newcommand{\lett}{}
\newcommand{\spage}{}
\newcommand{\rdate}{2000 May 17}
\newcommand{\adate}{2000 September 12}

\begin{document}

\title{LMSA and High-Redshift Dusty Starburst Mergers}

\author{Kenji  {\sc Bekki} \\
{\it 
National Astronomical Observatory, Mitaka, Tokyo, 181-8588}\\
{\it E-mail: bekki@th.nao.ac.jp}  
\\[6pt]
Yasuhiro  {\sc Shioya}\\
{\it Astronomical Institute,
Tohoku University, Sendai, 980-8578}\\ 
}

\abst{
By using a new numerical code for deriving the spectral energy
distributions of galaxies,
we have investigated the time evolution of morphological properties,
the star formation rate, and the submillimeter flux at 850 $\mu$m in  
high-redshift ($z$) 
dusty starburst mergers with  mass ratio ($m_{2}$) of two disks 
ranging from 0.1 (minor merger) to  1.0 (major one). 
We found that the maximum star-formation rate,
the degree of dust extinction, and the 850  $\mu$m flux
are larger for mergers with larger $m_{2}$. 
The 850 $\mu$m flux 
from  mergers at 1.5 $\le z \le$ 3.0 
in the observer frame is found to be a few  mJy
for major merger cases,
and at most $\sim$ 100 $\mu$Jy for minor ones. 
This result suggests that only high-redshift major mergers
are now detected by SCUBA with
the current 850 $\mu m$  detection limit of a few mJy.  
These results imply that LMSA with the expected 
detection limit of the order
of 10 $\mu$Jy at  850 $\mu m$ can be used to study 
high-redshift mergers with variously different $m_2$,   
and thus provide an important clue to the formation 
of galaxies in a  high-redshift universe.
}

\kword{galaxies: formation --- galaxies: interactions --- 
 infrared: galaxies}

\maketitle
\thispagestyle{headings}

\section
{Introduction}

One of fruitful and remarkable  achievements of
recent optical, infrared,  and radio observations of distant galaxies
is that interstellar dust has been  found to affect the photometric and spectroscopic 
evolution of galaxies at high redshift (e.g., Meurer et al. 1997;
Steidel et al. 1999). 
For example, deep surveys in the
submillimeter regime with the
Sub-millimeter Common-User Bolometer Array
(SCUBA) (Holland et al. 1999) on the James Clerk Maxwell Telescope
(Smail et al. 1997;
Hughes et al. 1998; Barger et al. 1998;
Smail et al. 1998;  Lilly et al.  1999),
in the Mid/Far-Infrared with the
 Infrared Space Observatory (ISO) (e.g., Flores et al. 1999),
and in the radio with the Very Large Array (VLA)
(e.g., Richards  1999)
have revealed the nature of starburst galaxies
whose star formation is heavily obscured by dust at intermediate
and high redshifts. 
These observations have simultaneously
demonstrated that optical data alone
can provide a partial view of the galactic evolutionary history,
which is possibly  qualitatively incorrect.

A major remaining  challenge is thus to
reveal how heavily distant  starburst galaxies are 
obscured by dust 
and what physical process determines
the degree of dust extinction in these galaxies. 
Galaxy merging has been  observed to be closely associated 
not only  with  
low-z dust starburst galaxies,  such as  ultra-luminous infrared galaxies
(Sanders, Mirabel 1996),  but also with intermediate/high-$z$ 
ones,  such as faint SCUBA sources (Smail et al. 1998). 
Accordingly,  it is primarily important
to investigate how intermediate/high-$z$ galaxy merging can
determine the nature of dusty starburst galaxies.
Bekki, Shioya, and  Tanaka (1999) first investigated
both the morphological and photometric evolution
of high-$z$ dusty starburst major mergers,
and found that major mergers can
successfully reproduce
the morphological and photometric 
properties of faint SCUBA
sources observed by Smail et al. (1998). 
However, they only described the evolution
of dusty starburst $major$ mergers and did not 
consider at all how dusty gas is important for
the evolution of  minor mergers with the mass ratio
of two merging disks less than 0.1 ($m_2$ $<$ 0.1)
and that of unequal-mass ones with $m_2$ $\sim$ 0.3.
Minor merging is considered to be occurring  more frequently than major merging,
and is important for the growth of bulges (e.g., Mihos,  Hernquist 1994),
whereas an unequal-mass one has been demonstrated to be related to the 
formation of S0s (Bekki 1998). 
Thus, investigating different types of dusty mergers (with different $m_2$)
not only leads  to a  better understanding of the nature 
of dusty starburst galaxies,  
but also provides a new clue to the origin of the Hubble sequence.

This paper considers  how
the mass ratio,  $m_2$,
controls  the time evolution of the morphological properties,
the degree of dust extinction, and the  submillimeter flux at  850 $\mu$m 
in dusty starburst mergers.
We furthermore demonstrate how the 850 $\mu$m flux 
in the observer-frame depends on $m_2$ if mergers are located
at $z$ = 1.5 and 3.0.
Based on our present numerical results, we discuss
whether the Large Millimeter and Submillimeter
Array (LMSA) can detect reemission of dusty gas at 850 $\mu$,
and suggest that LMSA can particularly play a vital role in revealing
morphological transformation processes of galaxies
in a  high-redshift universe.  

\section
{Model}

  We here consider both the time  evolution of the galactic morphology
and that of the spectral energy distributions (SEDs),
based on the results of numerical simulations that could  follow
both  the dynamical and chemical evolution of galaxies.
The numerical techniques for solving galactic
chemodynamical and photometric evolution
and the methods for deriving SEDs in numerical
simulations of galaxy mergers with dusty starburst
are given in  Bekki and  Shioya (2000).
The advantages and disadvantages of the model in reproducing
the observed SED of ULIRGs are summarized in detail by Bekki and  Shioya (2000).
For example, our model does not include physical processes
related to dust production, grain destruction, and grain growth 
in the interstellar medium,  and thus can not follow the time evolution
of dusty starbursts from the optically thick dusty cocoon stage
to the optical thin one  in detail. Mazarellar et al. (1991) compared
the far-IR properties of 187 Markarian galaxies with those of the IRAS Bright
Galaxy Sample and  extensively discussed  the nature of these in terms
of the grain-size distribution. Although our model can not include  this
important  effect of the grain-size distribution (and dust compositions)
on the photometric properties of dusty starburst galaxies, 
owing to the adopted simple model assumption in the present paper,
we should investigate this in the future.

  We constructed  models of galaxy mergers between two bulgeless gas-rich disks
  embedded in  massive dark-matter
  halos by using the Fall-Efstathiou  model (Fall and  Efstathiou 1980).
  The initial mass-ratio of dark matter halo
  to disk was  4:1 for the two  disks.
  The mass ratio of the smaller disk (referred to as the `secondary') to the larger one
  (the `primary') in a merger, which is represented by $m_{2}$,
  was considered to be a critical determinant
  for the evolution of dusty starburst mergers in the present study.
  We mainly describe the results of the model with $m_{2}$ = 0.1 
(minor merging), 0.3 (unequal-mass), 0.5, and 1.0 (major).
The disk mass ($M_{\rm d}$) was 
6.0 $\times$ $10^{10}$ $ M_{\odot}$ for the primary
in all merger models of the present study.
The  exponential disk scale length  
and the maximum rotational velocity for disks were 
$3.5 {(M_{\rm d}/6.0 \times 10^{10}   M_{\odot})}^{1/2} $ kpc
and $ 220  {(M_{\rm d}/6.0 \times 10^{10} \rm M_{\odot})}^{1/4}$ km ${\rm s}^{-1}$,
respectively.
These scaling relations were 
 adopted so that
 both the Freeman law and the Tully Fisher relation with
 a constant mass-to-light ratio could  be satisfied for the
structure and kinematics of the two disks. 
 For example, parameter values for disk structure 
and kinematics for the model with $m_{2}=0.3$ were as follows.
 The size and mass of  a disk are set to be 
 17.5 (9.6) kpc and 6.0 (1.8)$\times 10^{10}  M_{\odot}$, respectively,
 for the primary (the secondary).
 The scale length and the scale height of an initial exponential disk is
 3.5 (1.9) kpc and 0.7 (0.38) kpc, respectively,  for the primary (the secondary). 
 The rotational curve of the primary (the secondary)
 becomes flat at 6.1 (3.4) kpc with the maximum velocity of 220 (163) km/s.
 The Toomre stability parameter (Binney, Tremaine 1987) for initial disks was
 set to be 1.2.

The collisional and dissipative nature of disk interstellar gas was represented by
discrete gas (Schwarz 1981) and the initial gas mass fraction
was set to be 0.5 (corresponding to a very gas-rich disk).
The mass and the size for each of the clouds were 
3.0 $\times$ $10^6$ $M_{\odot}$ and 130 pc, respectively.
 Star formation in gas clouds during galaxy merging is modeled by converting gas particles
 to stellar ones according to the Schmidt law with  an  exponent
 of 2.0 (Schmidt 1959; Kennicutt 1989).
Although the effects of supernovae feedback on dynamical evolution
of the mergers are not included, the effects
probably would  not change significantly the present numerical results,
mainly because the adopted mass of a merger progenitor disk was 
fairly large. 
 Stellar components that are originally gaseous  ones are referred to as new stars. 
% The present numerical results  do not depend so strongly on the exponent
% in the Schmidt law.
% The thermal and dynamical effects of supernovae feedback  
% are not included, however, such effects probably modify the present 
% results only slightly.
 Chemical enrichment through star formation during galaxy merging
was assumed to proceed both locally and instantaneously in the present study.
The fraction of gas returned to interstellar medium in each stellar particle
and the chemical yield
were 0.3 and 0.02, respectively.
The initial metallicity $Z_{\ast}$ for each stellar and gaseous
particle in a given galactic radius  $R$  (kpc) from the center
of a disk was given 
according to the observed relation $Z_{\ast} = 0.06  {10}^{-0.197 \times (R/3.5)}$
of typical late-type disk galaxies (e.g., Zaritsky et al. 1994).
% The intrinsic spin vector of the primary is exactly parallel with $z$ axis
% whereas that of  the secondary is tilted by ${30}^{\circ}$ from $z$ axis.
% The orbital plane of the merger is exactly the same as the  $xy$ plane.
%The orbit of a galaxy merger is assumed to be parabolic
%and its pericenter distance and initial separation of the two disks  
%are set to be 17.5 kpc and 140 kpc, respectively.
%Total particle number is 20000 for dark matter, 20000 for old stellar 
%components (initially located within disks), and 20000 for gaseous ones
%For a major merger model.
All of the simulations were carried out on the GRAPE board (Sugimoto et al. 1990)
with a 
gravitational softening length of 0.53 kpc.
For calculating the SED
of a merger, 
we use the 
spectral library GISSEL96, 
which is the  latest version of  Bruzual and  Charlot (1993).
The SEDs of dusty mergers and the way to calculate
the 850 $\mu$m flux from  high redshift mergers
both for  observer-frame and for rest-frame   are described in detail by
Bekki et al. (1999) and Bekki and  Shioya (2000).
 In the following, the cosmological parameters
 $ H_0$ and $ q_0$ are  assumed  to be 50 km s$^{-1}$ Mpc$^{-1}$ and 0.5,
 respectively.

% Fig.1

\section
{Results}

Figure 1 gives the  $m_2$ dependence of the morphological
properties of mergers at the epoch of maximum starburst.
For the  minor merger model with $m_2$ = 0.1, the secondary
sinks into the deep gravitational potential well
of the primary owing to the strong dynamical friction, 
and is finally  completely destroyed 
 in the central part of the primary.
The primary in the merger, on the other hand,
leaves  its disk morphology even after merging, 
though the primary suffers from dynamical
heating due to the accretion of the secondary, and consequently
forms a thick
disk.
Furthermore,  mergers with larger $m_2$ become more similar to
early-type galaxies (E/S0), principally because the disk destruction
due to tidal gravitational force and dynamical relaxation
more drastically  proceeds  for mergers with larger $m_2$. 
These results suggest that 
$m_2$  is an important factor for controlling the morphological
properties of dusty starburst galaxies formed by merging. 
These  results also imply that
$m_2$ is one of important factors for the origin
of the Hubble sequence.

Figure 2 shows how  
the star-formation rate and the degree of dust extinction
at the epoch of maximum starburst in mergers
depend on $m_2$.
Here, we estimate $A_{V}$  for a given $m_2$ by comparing 
the $B-V$ color in the model with dust extinction
and that in the model without dust extinction.  
The maximum star-formation rate is larger for mergers with
larger $m_2$, essentially because a larger amount of interstellar
gas can be transferred to the central region, and consequently converted
into new stars in mergers with larger $m_2$.
Furthermore, $A_{V}$ is larger for mergers with
larger  $m_2$, which means that 
young luminous  populations formed by starburst
are more heavily obscured by dusty gas for larger  $m_2$.
For mergers with larger  $m_2$,
the stronger tidal disturbance triggers
efficient cloud--cloud collisions, and consequently
induces a larger amount of gaseous dissipation.
As a natural result,
the dusty interstellar gas 
is more efficiently transferred to the central regions
to form  higher  density gaseous regions surrounding
the nuclear starburst populations in mergers with larger $m_2$.  
This is the main reason for the above $m_2$ dependence
of $A_{V}$.

Figure 3 shows the following three new and important results on
the dependence of the observed 850 $\mu$m flux from
high-redshift mergers on
the redshift and $m_2$.
Firstly, the observed 850 $\mu$m flux is not so significantly
different between dusty starburst mergers at  $z$ = 1.5 and 2.0
owing to a large negative $K$-correction in the submillimeter range,
though the flux is smaller in higher redshift mergers 
for a given $m_2$. 
Secondly, the observed 850 $\mu$m flux is larger for mergers with
a larger  $m_2$. 
This is firstly because total amount of stellar light from
young massive stars, which can be reemitted in dusty gas,
is larger for mergers with the larger $m_2$,
and secondly because a larger amount of dusty gas
can be transferred to the central region,  and consequently obscure
more heavily starburst populations in mergers with larger $m_2$.
Thirdly,  the observed 850 $\mu$m flux greatly  exceeds the current SCUBA detection
limit of $\sim$ a few mJy for larger $m_2$ (= 0.5, 1.0).
Considering the results shown in figure 1,
it is  strongly suggested  that high-redshift  dusty early-type galaxies
which are being formed by mergers with larger $m_2$  are preferentially
detected by SCUBA.
Furthermore, this result implies that
dusty starburst triggered 
by minor merging, which is considered to be  more frequently 
occurred than major merging with $m_2$ = 1.0 
and an important determinant for the dynamical evolution of early-type
disk galaxies (e.g., Mihos,  Hernquist 1994),
should  not be studied   by SCUBA,  but by a future  submillimeter array
with a  detection limit of 10 $-$ 100 $\mu$Jy.

% Fig.2
% Fig.3

\section{Discussion and Conclusion}

Considering the present numerical results,
LMSA has the following three advantages in
studying the origin and the nature of high redshift  dusty  
starburst galaxies.
Firstly, LMSA is expected to 
detect a submillimeter flux of 10 $-$ 100 $\mu$Jy at 850 $\mu$m
and therefore  can investigate not only high-redshift 
dusty starburst galaxies
formed by  major mergers, but also those by minor and unequal-mass ones.  
Since minor and unequal-mass mergers are suggested to be
important for the formation of Sa and for that of S0
(Mihos, Hernquist 1994; Bekki 1998),  respectively,
extensive studies of  dusty galaxies with   850 $\mu$m
flux of the order of 10 $-$ 100 $\mu$Jy  by  LMSA
will  provide a new clue to the origin of the Hubble sequence.
Secondly, LMSA can investigate the nature of dusty starburst
galaxies at  variously different dynamical stages of merging.
Bekki and Shioya (2000) have demonstrated that
the 850 $\mu$m flux from dusty starburst populations
depends strongly on  the degree of the dynamical
relaxation of merging galaxies.
Accordingly,   LMSA will  find a  physical relationship
between the dynamical evolution and the photometric one at the submillimeter band
in high-redshift dusty starburst galaxies,
if future deep optical and near-infrared morphological,
structural, and kinematical studies will have revealed the dynamical
conditions of these galaxies.

Thirdly,  and  most importantly, LMSA  
can provide valuable information concerning  forming disk galaxies at high redshift. 
Recent numerical simulations based on the hierarchical
assembly of cold dark-matter (CDM) halos 
have demonstrated that the star formation rate
is estimated to be on the  order of $\sim$ 10 $M_{\odot}$ ${\rm yr}^{-1}$
when successive minor merging of subgalactic  
clumps becomes important for the formation of galactic disks 
at z = 1 $-$ 2 (e.g., Steinmetz,  M\"uller 1995;
Bekki, Chiba 2000).
The present study has
demonstrated that 
the 850 $\mu$m flux in minor mergers at the redshift of 1 $-$ 3
is at most on the  order of  $\sim$ 100 $\mu$Jy
and,  accordingly,  high redshift forming disk galaxies
can be investigated by  LMSA rather than by 
SCUBA.
We suggest that   
a  detailed  comparison between LMSA submillimeter studies
and future optical and infrared spectrophotometric ones 
would  reveal  
the degree of dust extinction ($A_V$)
and the physical parameters for $A_V$ in forming disk galaxies.

By using a new and original code for calculating the SEDs
of dusty starburst galaxies, we have demonstrated 
that  the mass ratio,  $m_2$, controlling
the strength of  tidal disturbance 
(or the degree of disk destruction) 
in galaxy merging
is an important determinant 
for the maximum star-formation rate and  the degree of  
dust extinction ($A_V$).
The derived dependence on  $m_2$  does not depend
very  strongly on other  parameters, such as
the orbital configurations, internal structure, and gas mass 
fraction of galaxy mergers  (Shioya, Bekki,  in preparation).
Based on these  essentially new results,
we emphasize the importance of LMSA in investigating and revealing
the nature of high-redshift dusty starburst galaxies.
We lastly expect that future observational studies of  
high-redshift galaxy mergers with different  $m_2$  will provide
new clues to the formation  and  evolution of galaxies.

\par
\vspace{1pc}\par
Y. S. thanks to  the Japan Society for Promotion of Science (JSPS)
Research Fellowships for Young Scientist.

\section*{References}
\small

%\bibitem[Barger et al.  1998]{ba98}
\re
Barger A. J., Cowie L. L., Sanders D. B., Fulton E., Taniguchi Y.,
Sato Y., Kawara K., Okuda H. 1998, Nature 394, 248

%\bibitem[Barger et al.  1999]{ba99}
%Barger A. J., Cowie, L. L., Smail, I., Ivison, R. J., Blain, A. W., Kneib, J-P.
%1999, astro-ph/9903142

\re
Bekki K., 1998, ApJ 502, L133

\re
Bekki K., \& Chiba M. 2000, ApJ 534, L89 

\re
Bekki K., \& Shioya Y. 2000, ApJ in press 

\re
Bekki K., Shioya Y., Tanaka I.  1999,  ApJ, 520, L99

%\bibitem[Binney \& Tremaine 1987]{bt87}
\re
Binney J.,  Tremaine S. 1987,  Galactic Dynamics, (Princeton
Univ. Press, Princeton) page 313.

\re
Bruzual A. G.,  Charlot S. 1993, ApJ 405, 538

%\bibitem[Fall \& Efstathiou 1980]{fe80}
\re
Fall S. M.,  Efstathiou G. 1980, MNRAS 193, 189

\re
Flores H., Hammer F., Thuan T. X., C\'esarsky C.,
Desert F. X., Omont A., Lilly S. J., Eales S.,
Crampton D., \& Le F\'evre O. 1999, ApJ 517, 148

%\bibitem[Holland et al. 1999]{ho99}
\re
Holland W. S., Robson E. I., Gear W. K., Cunningham C. R.,
Cightfoot J. F., Jenness, T., Ivison R. I., Stevens J. A. et al.  1999, MNRAS, 303, 659

%\bibitem[Hughes et al. 1998]{hug98}
\re
Hughes, D. H., Serjeant S., Dunlop J., Rowan-Robinson H., Blain A.,
Mann R. G., Ivison R., Peacock J. et al., 1998, Nature 394, 241

%\bibitem[Kennicutt 1989]{ken89}
\re
Kennicutt R. C. 1989, ApJ 344, 685

%\bibitem[Lilly et al. 1999]{li99}
\re
Lilly, S. J., Eales, S. A., Gear, W.K.P., Hammer, F., Le F\'evre, O.,
Crampton D., Bond J. R., Dunne L. 1999, ApJ 518, L641

\re
Mazzarella, J. M., Bothun, G. D.,  Boroson, T. A.
1991, AJ 101, 2034 

\re
Meurer G. R, Heckman T. M., Lehnert M. D., Leitherer C., 
Lowenthal J. 1997, AJ 114, 54

%\bibitem[Mihos \& Hernquist 1994]{mh94}
\re
Mihos J. C.,  Hernquist L. 1994, ApJ 425, L13

%\bibitem[Mihos \& Hernquist 1996]{mh96}
%Mihos, J. C., \& Hernquist, L. 1996, \apj, 464, 641 

%\bibitem[Richards 1999]{ri99}
\re
Richards E. A., 1999, ApJ 513, L9

%\re
%Sakamoto, S. 2000, private communication

\re
Sandage A. 1961, The Hubble Atlas of Galaxies (Washington, DC: Carnegie Inst.
Washington)

%\bibitem[Sanders et al. 1988]{ss88}
%Sanders, D. B., Soifer, B. T., Elias, J. H., Modore, B. F.,
%Mattehews, K., Neugbauer, G., Scoville, N. Z. 1988a, \apj, 325, 74

%\bibitem[Sanders et al. 1988b]{ss88b}
%Sanders, D. B., Scoville, N. Z., Sargent, A. I., \& Soifer, B. T.
%1988b, \apj, 324, L55 

%\bibitem[Sanders \& Mirabel 1996]{sm96}
\re
Sanders D. B.,  Mirabel I. F. 1996 ARA\&A, 34, 749

%\bibitem[Schmidt 1959]{sch}
\re
Schmidt M. 1959, ApJ 344, 685

%\bibitem[Schwarz 1981]{sch81}
\re
Schwarz M. P. 1981, ApJ 247, 77

%\bibitem[Smail  et al. 1997]{sm97}
\re
Smail I.,  Ivison R. J.,   Blain A. W. 1997, ApJ 490, L5 

%\bibitem[Smail  et al. 1998]{sm98}
\re
Smail I.,  Ivison R. J.,   Blain A. W.,   Kneib J.-P. 1998, 
ApJ, 507, L21

\re
Steidel C. C., Adelberger K. L., Giavalisco M.,
Dickinson M.,  Pettini M. 1999, ApJ 519, 1

\re
Steinmetz M.  M\"uller E. 1995, MNRAS 276, 549

%\bibitem[Sugimoto et al. 1990]{sug90}
\re
Sugimoto, D., Chikada, Y., Makino, J., Ito, T., Ebisuzaki, T., 
Umemura, M. 1990, Nature 345, 33

%\bibitem[Toomre \& Toomre 1972]{tt72}
%Toomre, A., \& Toomre, J. 1972, \apj, 178,623

%\bibitem[Toomre 1977]{too77}
%Toomre, A. 1977, in  The evolution of galaxies and stellar Populations 
%ed. by B. Tinsley \& R. Larson (New. Haven. CN: Yale Univ. Press), p401

%\bibitem[Wielen 1977]{wie77}
%Wielen, R. 1977, \aap, 60, 263

%\bibitem[Wright et al. 1990]{wr90}
%Wright, G. S., James, P. A., \& Joseph, R. D., \& Mclean, I. S.
%1990, \nat, 344, 417

%\bibitem[Zaritsky et al. 1994]{za94}
\re
Zaritsky D., Kennicutt R. C. Jr, Huchra J. P. 1994, ApJ 420, 87

\label{last}

\newpage

\section*{Figure Captions}

% Fig.1
\begin{fv}{1}{18pc}%
{Dependence of the morphological properties  of mergers 
at the epoch of the maximum star-formation rate 
projected 
onto the $xz$ plane on the mass ratio of two merging disks ($m_2$).
Stars, gas, and new stars are plotted
in the same panel.
We here do not display dark halo components in order to  show more clearly
the morphological properties  of the disk components.
Each frame measures 77 kpc, and the mass ratio of  $m_2$
is indicated in the upper-left
corner of each panel.
The epoch of maximum star-formation rate is $T$ = 3.06 Gyr
for $m_2$ = 0.1, 1.63 Gyr for $m_2$ = 0.3, 1.50 Gyr for $m_2$ = 0.5
and 1.29 for $m_2$ = 1.0,  
where $T$ represents  the time
that has elapsed since the two disks begin to merge for each of
the four models. 
Note that the primary disk of the merger with larger $m_2$ 
is more greatly destroyed by the accretion of the secondary.
The final morphology of each of merger remnant in nearly
virial equilibrium ($\sim$ 1 Gyr after maximum starburst)
is basically similar to that at the epoch of maximum starburst.   
The final morphology also depends  on $m_2$,  such that
the merger remnant with larger $m_2$ is 
more similar to early-type galaxies (E/S0). 
We stress that the  merger remnant with $m_2$ = 0.3
is more similar to S0s with no remarkable thin stellar disk
such as NGC 3245 and 4684 than those with thin extended stellar disks such
as NGC 4111 and 4710 [see the Hubble Atlas of Galaxies of  Sandage (1961)]}.
\end{fv}

% Fig.2
\begin{fv}{2}{18pc}%
{The $m_2$ dependence of maximum star formation rate in units of 
$M_{\odot}$ ${\rm yr}^{-1}$ (upper panel) and 
that of the degree of dust extinction 
($A_{V}$)
in units of mag (lower one).
Note that both the maximum star-formation rate and the degree
of dust extinction is larger for dusty starburst mergers with
a larger $m_2$. 
Also note that the maximum star-formation
rate in the major merger with $m_2$ = 1.0 is $ \sim$   378  
$M_{\odot}$ ${\rm yr}^{-1}$, which
corresponds roughly to (or is larger than)
that required for explaining infrared luminosity in 
ultra-luminous infrared galaxies (Sanders, Mirabel 1996)}.
\end{fv}

% Fig.3
\begin{fv}{3}{18pc}%
{The $m_2$ dependences of 850 $\mu$m flux of high redshift mergers
at $z$ = 1.5 (solid line) and 3.0 (dotted one) in the observer-frame.
The current detection limit in SCUBA (2 mJy) 
and the expected one 
in LMSA (40 $\mu$Jy) are also given  by the thick  solid lines.
Note that in comparison  with SCUBA, 
LMSA can detect not only dust reemission from major mergers at high $z$
but also that from minor ones.}
\end{fv}

\end{document}